\documentclass[12pt]{iopart}  
\usepackage{graphicx}  
\usepackage{textcomp, cite}
\usepackage{color}
\usepackage[utf8]{inputenc} 
\newcommand{\xvector}[1]{\mathbf{x}_{\mathrm{#1}}} 

\begin{document}

\title
{A three-dimensional self-learning kinetic Monte Carlo model: application to Ag(111)}
\date{\today}
\author{A. Latz, L. Brendel and D.E. Wolf}
\address{Department of Physics and Center for Nanointegration Duisburg-Essen (CeNIDE), University of Duisburg-Essen, D-47057 Duisburg, Germany}
\ead{andreas.latz@uni-due.de}

\begin{abstract}
The reliability of kinetic Monte Carlo (KMC) simulations depends on accurate transition rates.
The self-learning KMC method (Trushin \textit{et al.}, \textit{Phys. Rev.} \textbf{B} 72, 115401 (2005)) combines the accuracy of rates calculated from a realistic potential with the efficiency of a rate catalog, using a pattern recognition scheme.
This work expands the original two dimensional method to three dimensions.
The concomitant huge increase in the number of rate calculations on the fly needed can be avoided by setting up an initial database, containing exact activation energies calculated for processes gathered from a simpler KMC model.
To provide two representative examples, the model is applied to the diffusion of Ag monolayer islands on Ag(111), and the homoepitaxial growth of Ag on Ag(111) at low temperatures.
\end{abstract}


\maketitle

\section{Introduction}
In many fields of solid state physics, knowledge of the dynamics on the atomic scale is essential for understanding macroscopic phenomena.
Two examples are the complex morphological structures growing in epitaxy experiments\cite{Evans2006, Cox2005, Li2008, Li2009, Sindermann2011} and the electromigration induced failure of interconnects\cite{Sorbello1997, Tan2007, Tu2007,  Latz2012, Latz2012MRS}.
Typical experiments last from minutes to days.
To understand the macroscopic observations on a microscopic level, atomistic simulations can provide valuable insights.
Two common approaches are mostly used for atomistic simulations, molecular dynamics (MD) and kinetic Monte Carlo (KMC) simulations\cite{Bortz1975, Voter1986, Voter2007}.
In MD simulations the system's trajectory is integrated in time.
The time increment needed to resolve atomic vibrations limits the achievable time scales to ms.
KMC simulations, on the other hand, can reach much longer time scales.
Instead of following the system's trajectory, transitions between adjacent minima in the potential energy surface (PES) are directly performed.
The next process is chosen proportional to its transition rate.
If all of these processes are included with their correct rate, the statistics of the KMC trajectories will be indistinguishable from the one of the MD trajectories on scales larger than the lattice constant.

Since the transition rates strongly depend on the current system configuration, building up a correct rate catalog is a challenging task.
Thus, in most KMC models simple approximations are used for choosing processes to be considered, and calculating their rates.
Recently there has been substantial progress regarding the completeness issue of the rate catalog.
At each state, saddle point searches are performed to find possible transitions\cite{Henkelman2001}.
Trushin \textit{et al.}\cite{Trushin2005, Shah2012} combined this approach for two dimensional lattice models with a pattern recognition scheme.
In this so called self-learning kinetic Monte Carlo (SLKMC) method, previously calculated rates can be reused.
If the average number of  rates calculated on the fly per KMC step is small, the performance of the SLKMC method is as good as that of simpler approaches.
There has also been progress in the development of off lattice SLKMC models\cite{Kara2009, Nandipati2012}.
But the vast number of occurring processes allows only the simulation of rather small systems.

Up to now the SLKMC method was not expanded to three dimensions.
An expansion increases the number of possible transitions tremendously, which complicates an efficient simulation.
However, we found a way to apply the SLKMC method to three dimensional systems.
The number of rates calculated on the fly was reduced significantly by setting up an initial database.
It contains the exact activation energies for a set of processes, which were gathered from an auxiliary simulation using crude, but simple approximate activation energies.
This enabled us to simulate large systems without a significant overhead.
We demonstrate the versatility and performance of our model in two examples, the diffusion of large monolayer islands and the homoepitaxy of Ag on Ag(111).

The paper is organized as follows.
In the next three sections, details of our  three dimensional SLKMC simulation model are presented.
Section \ref{sec:precalculation} includes the details of the rate pre-calculation.
The two simulation examples are presented in section \ref{sec:Simulation}.
Finally, section \ref{sec:Conclusion} contains concluding remarks.

\section{KMC Simulation Model\label{sec:Model}}
In this work three dimensional Ag clusters on Ag(111) surfaces are investigated.
Our model is general enough, however, to be applicable to other surfaces as well.
The dynamical evolution of the Ag atoms is simulated by a sequence of thermally activated hopping processes, using a standard KMC algorithm \cite{Bortz1975, Voter1986, Voter2007}.
Atoms are restricted to fcc sites.
Single atoms are allowed to hop to empty nearest- and second nearest-neighbor fcc sites.
Due to the restriction to fcc sites, the later processes are necessary, as will be shown below.
The restriction to single atom processes is valid, since only large atom clusters, where island diffusion takes place mainly via edge diffusion of single atoms\cite{Karim2011}, are simulated.
For small clusters, the inclusion of concerted moves becomes important\cite{Karim2011}, which can be included in future work.
Exchange processes, where an atom exchanges position with another atom, which continues the diffusion, are neglected, because their activation energies are either comparable or higher than those of the corresponding hopping processes\cite{Ala-Nissila2002}.

\begin{figure}[tb]
\includegraphics[width=0.5 \linewidth]{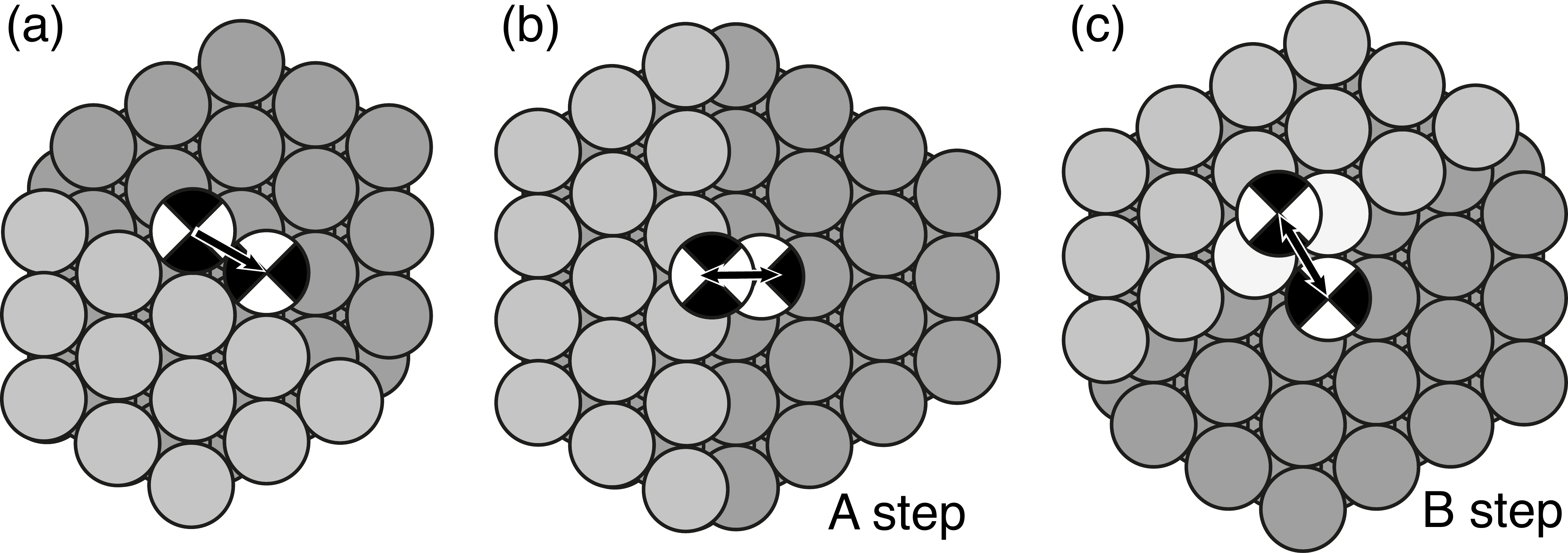}
\centering
\caption{Examples for the allowed hopping processes.
The black/white circles represent initial and final hopping positions.
(c):  Hops to empty second nearest-neighbor sites are only allowed, if the initial and final hopping positions are separated by a step edge consisting of two adjacent atoms, like the light gray ones.
\label{fig:allowed_hops}}
\end{figure}

For three dimensional Ag clusters. three kinds of hopping processes are distinguished.
Examples are shown in figure~\ref{fig:allowed_hops}:
Hopping processes to empty nearest-neighbor sites in the same surface plane (Fig.~\ref{fig:allowed_hops}(a)),
hopping processes to empty nearest-neighbor sites in the surface plane above or below (Fig.~\ref{fig:allowed_hops}(b)) and
hopping processes to empty second nearest-neighbor sites, which for (111) surfaces are always in a surface plane above or below (Fig.~\ref{fig:allowed_hops}(c)).
Some fcc sites cannot represent a minimum in the PES (e.g. left marked site in Fig.~\ref{fig:allowed_hops}(b)).
They are (unstable) saddle points that need to be crossed to reach a neighboring stable site.
To keep the possibility to reach those sites, also hops to empty nearest neighbor sites with two occupied nearest-neighbors ($Z=2$) are allowed.
Thus, some diffusion processes are divided into two steps in our model. 

On an fcc (111) surface, two step orientations are possible, so called A (Fig.~\ref{fig:allowed_hops}(b)) and B steps (Fig.~\ref{fig:allowed_hops}(c)).
Since the saddle point for crossing the latter is close to an hcp site, crossing a B step is impossible if only hops to nearest-neighbor sites with $Z\geq 2$ are possible\cite{Latz2012}.
To incorporate these important processes, hops to empty second nearest-neighbor sites with $Z\geq 2$ are allowed if the initial and final hopping positions are separated by a step edge consisting of two adjacent atoms, like the light gray ones in figure~\ref{fig:allowed_hops}(c).

The hopping rates of the allowed processes are given by the Arrhenius law
\begin{equation}
\nu = \nu_0 \exp\left( -\frac{E_\mathrm a}{k_\mathrm B T}\right)\ ,
\label{eq:RateCalculation}
\end{equation}
where $\nu_0$ denotes the attempt frequency and $E_\mathrm a$ is the
activation energy barrier, separating the initial and final configuration in the PES of the system.
 $k_{\mathrm {B}}$ is the Boltzmann constant and $T$ is the temperature.
A commonly used constant value\cite{Voter2007}  of $\nu_0=10^{12}~\mathrm{s}^{-1}$ is assumed.
To describe the interactions between Ag atoms, a many-body tight-binding-potential is used, which was fitted to experimental values of Ag by Cleri and Rosato\cite{Cleri1993}.

To calculate the rate for a transition, one needs to know the activation energy $E_{\mathrm a}$, which is the difference between the saddle-point energy 
$E_\mathrm b ^\mathrm s$ and the binding energy at the initial
fcc site, $E_\mathrm b^\mathrm i$.
Both energies strongly depend on the local environment of the hopping atom.
Thus, simplistic models for the activation energy can be very inaccurate for some processes.
Instead, whenever a local environment for an allowed transition appears for the first time, the corresponding activation energy is calculated.
These calculated activation energies are stored and can be reused by utilizing a pattern recognition scheme presented in the next section.

\section{Pattern Recognition\label{sec:Pattern}}
To save computation time, previously calculated activation energies are stored and can be reused utilizing a pattern recognition scheme.
Our pattern recognition scheme is a three dimensional expansion of the two dimensional one presented by Trushin \textit{et al}.\cite{Trushin2005}.

\begin{figure*}[tb]
\includegraphics[width=0.95\linewidth]{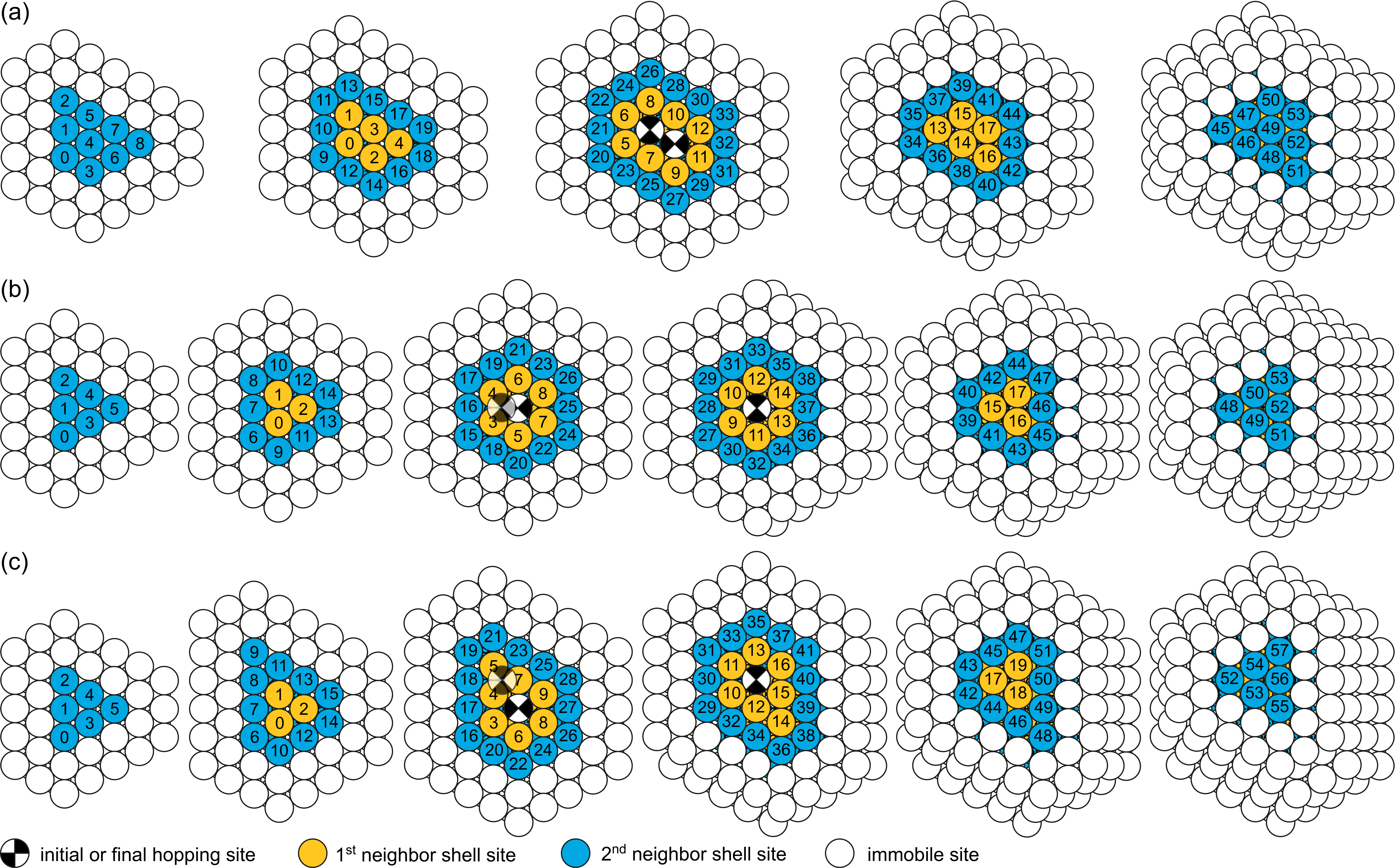}
\centering
\caption{
Two neighbor shells around the initial and final hopping site are used to identify a configuration.
(a)--(c) correspond to the three types of hopping processes defined in figure~\ref{fig:allowed_hops}.
(a): Hops to nearest-neighbor sites in the same surface plane, (b) hops to nearest-neighbor sites in a different surface plane, and (c) hops to second nearest-neighbor sites.
The three dimensional structure of the shells is made clear by stacking (111) layers as shown from the left to the right. 
\label{fig:PatternRecognition}}
\end{figure*}

In order to reduce the number of neighbors which have to be considered for the calculation of an activation energy, each hopping process is characterized by a limited number of neighbor sites of the initial and of the final site.
The nearest neighbors of these two sites form a first shell (see Fig.~\ref{fig:PatternRecognition}).
Further sites, up to fourth nearest-neighbors, form a second shell.
The atoms occupying sites in these two shells define what we call a configuration in the remainder.
For each of these configurations an activation energy is calculated (see section \ref{sec:AEC}).
Atoms further away have a relatively weak effect on the value of the activation energy, however they are necessary to stabilize the configurations.
For the purpose of calculating activation energies these atoms are put into fixed ideal fcc positions, which form a subset of the white  sites in figure~\ref{fig:PatternRecognition}. 

The white sites are themselves arranged in two shells, number three and four.
The third shell contains the nearest-neighbors of the second one, and the fourth the nearest-neighbors of the third one.
The occupied subset of sites in the third shell consists of all nearest-neighbors of occupied sites in the second shell within in the same layer or the layer above (see Fig.~\ref{fig:PatternRecognition}).
Likewise, the sites in the fourth shell are filled according to the occupation of the third shell within the same layer or the layer above.

Each configuration is identified by a unique combination of three indices to find its activation energy in a database.
To save memory and computation time, the symmetry of the fcc (111) surface is utilized:
Hops to nearest- or second nearest-neighbor sites (cf.\ Fig.~\ref{fig:allowed_hops}) are symmetry related to each other by a 120\textdegree\ or 240\textdegree\ rotation.
For hops within the same surface plane (e.g.\ Fig.~\ref{fig:allowed_hops}(a)) an additional reflection is possible.
A configuration is identified by a combination of three numbers.
The first number represents the hopping direction.
The second number is a bitwise representation of the occupancy of the first shell.
The occupancy of the second shell is coded in a third number, because the total number of possible configurations exceeds $2^{64}$, so that a single 64 bit word cannot store both shells together.
The sites of each shell  are numbered separately beginning with zero.
If the $i^{\mathrm{th}}$ site is occupied $2^i$ is added to the corresponding integer.
This triple of numbers serves as the key for the database, implemented as an unordered hash map\cite{Boost}.
Two example configurations can be found in figure~\ref{fig:Examples}.
In principle the pattern recognition scheme can be enhanced by also considering hcp sites.
The occupancy of the additional neighboring hcp sites would have to be encoded in additional bits.
In the next section, the activation energy calculation for the described transition configurations is presented.

\section{Activation energy calculation \label{sec:AEC}}
Each time a configuration occurs for the first time, its activation energy is calculated on the fly using the drag method\cite{Huang1991}.
For key processes, it showed only minor differences compared to the more sophisticated, but notably slower, nudged elastic band method\cite{Trushin2005}.
In the drag method, the hopping atom is dragged in discrete steps of $\Delta x$ from  $\xvector{i}$ to  $\xvector{f}$ along the hopping direction, defined by the vector connecting the ideal fcc sites.
At each step mobile atoms are allowed to relax to the local PES minimum using the Broyden-Fletcher-Goldfarb-Shanno algorithm (BFGS2 from the GNU Scientific Library)\cite{Galassi2009}.
The hopping atom relaxes under the constraint of a fixed coordinate along the hopping direction.
The maximum energy along the traveled path is the saddle point energy.
After each step, the neighbor atoms are set back to their initial fcc sites.
Thus, concerted moves are prohibited.
This is in line with our KMC model, which allows only single atom processes.
Since the saddle point energy is the same for the diffusion processes in both directions, each simulation yields two activation energies, namely for the forth and back process.
If the neighbor atoms would not be set back, the activation energies for concerted moves would be calculated.
To implement them in the KMC model, the key for the database would have to contain not only the initial configuration and the hopping direction, but a complete characterization of the final configuration.

\begin{figure*}[tb]
\centering
\includegraphics[height=2.8cm]{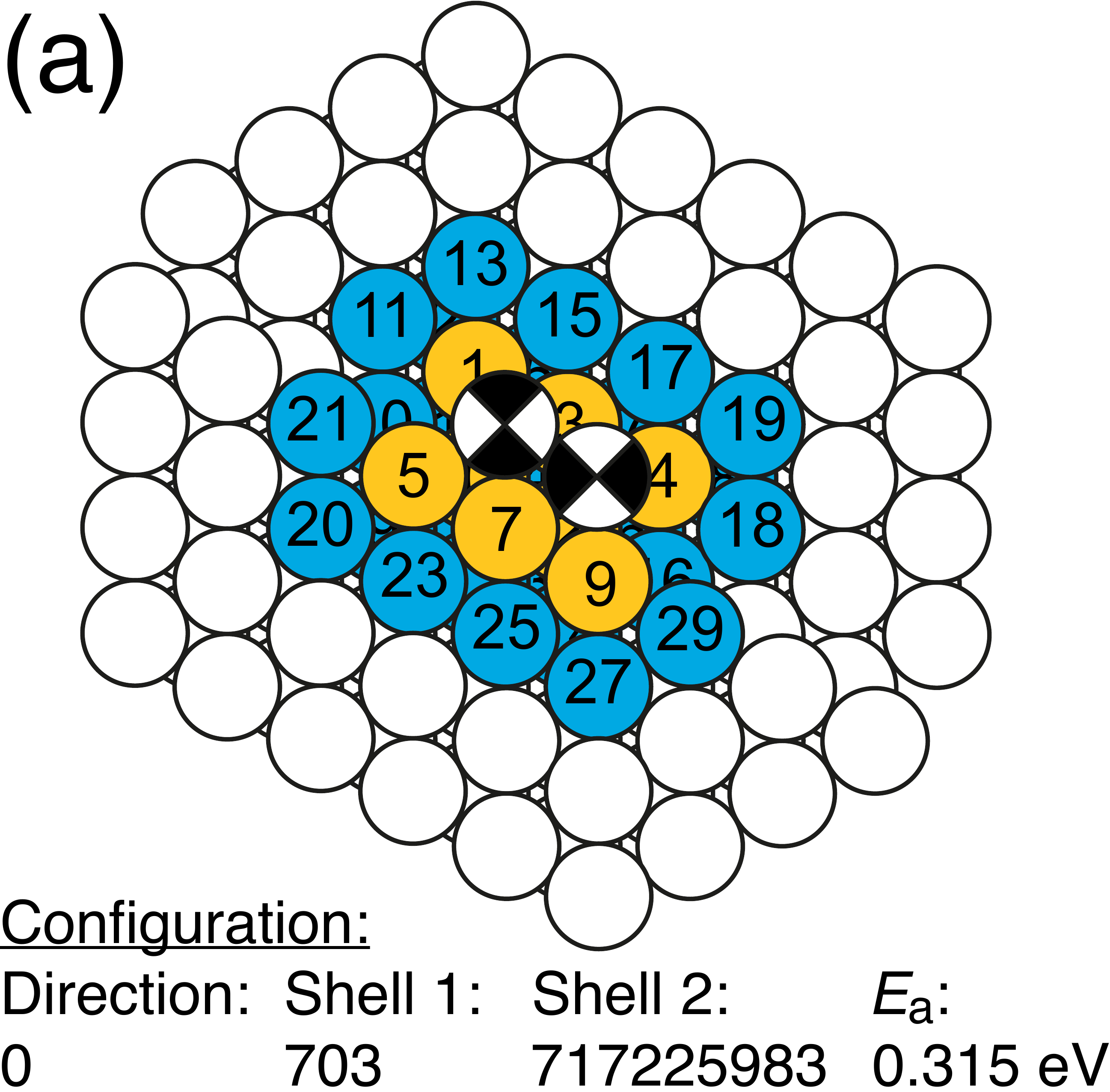}
\hspace{0.0cm}
\includegraphics[height=2.8cm]{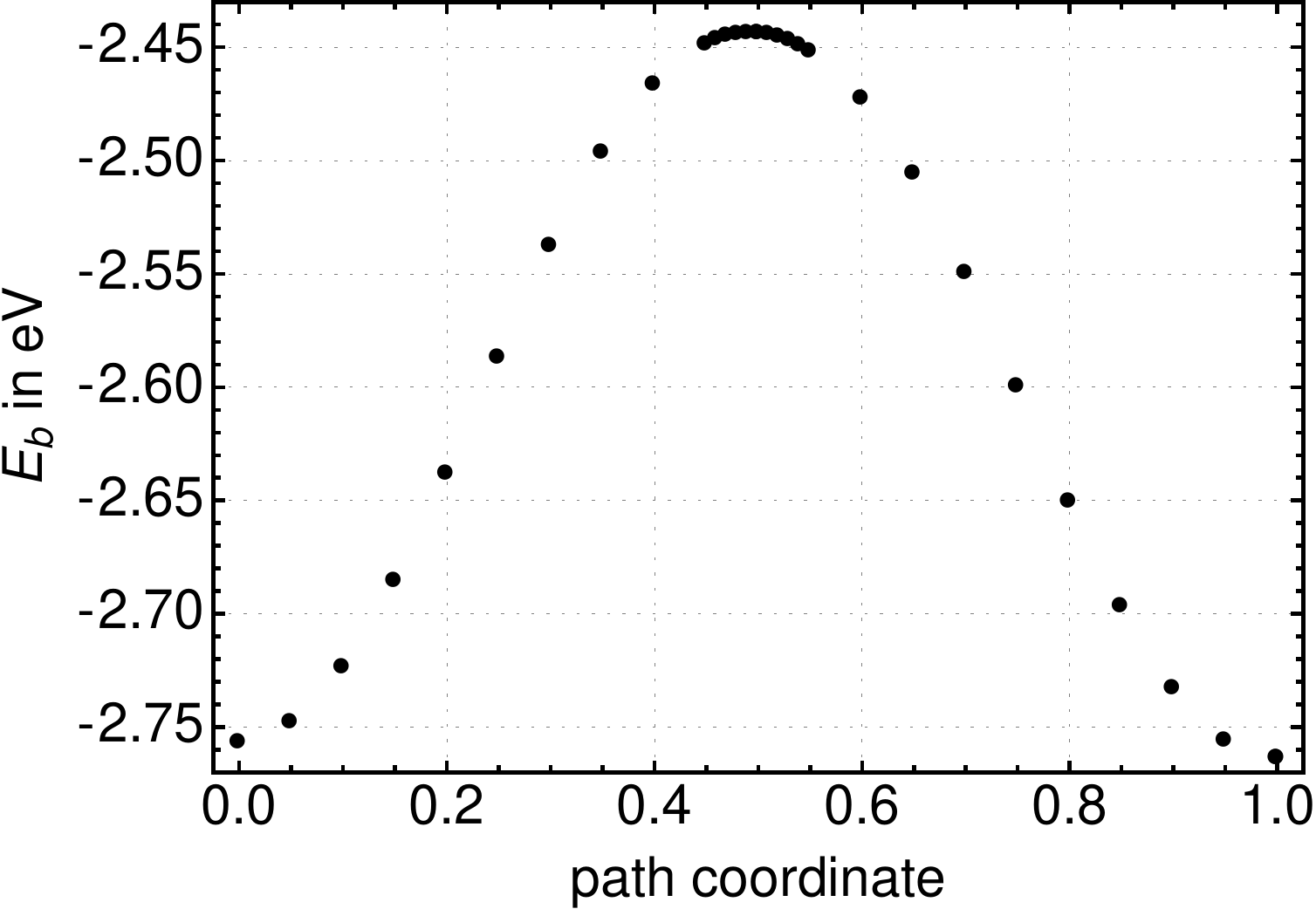}
\hspace{0.3cm}
\includegraphics[height=2.8cm]{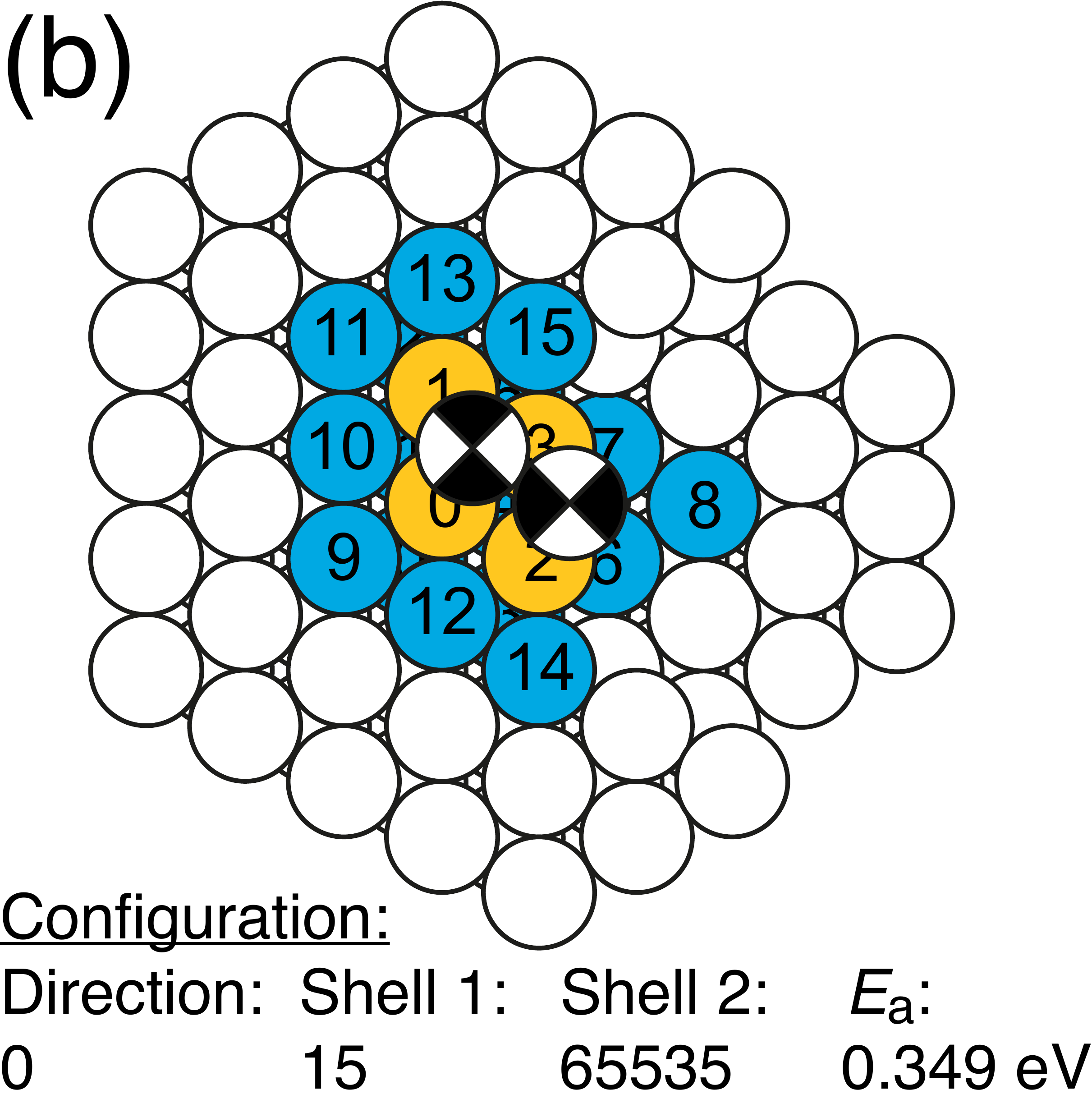}
\hspace{0.0cm}
\includegraphics[height=2.8cm]{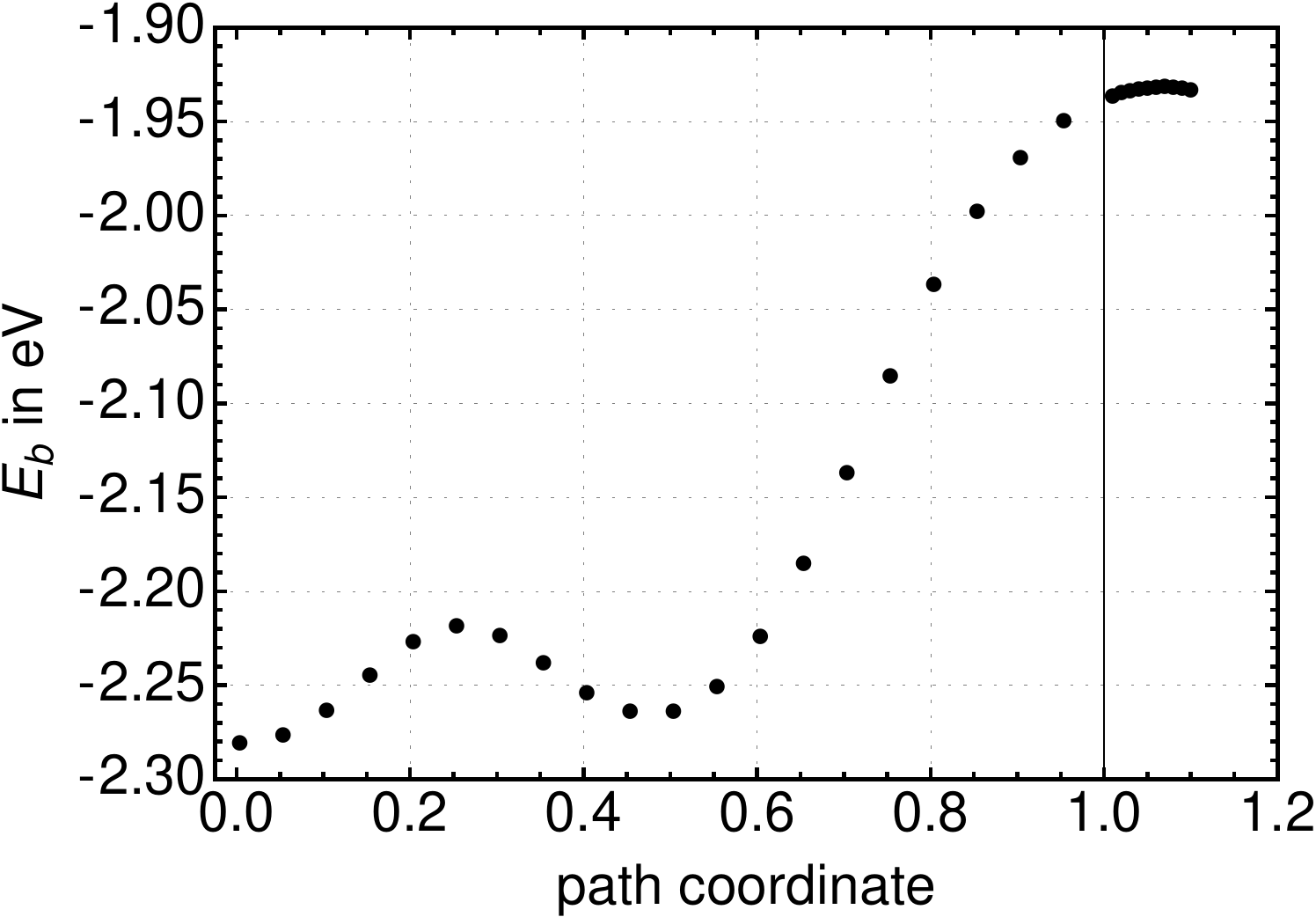}
\caption{Example configurations:
(a): Diffusion along step. 
The initial and final configuration are stable.
(b): Diffusion to a step edge.
The saddle point lies beyond the initial $x_{\mathrm{f}}$ (vertical line).
Thus the simulation interval was advanced.}
\label{fig:Examples}
\end{figure*}

Only the neighbor sites used for the pattern recognition (numbered in Fig.~\ref{fig:PatternRecognition}) are possibly occupied with mobile atoms.
To stabilize them, they are embedded in two shells of immobile sites  as described above.
Two example configurations are shown in figure~\ref{fig:Examples}.

To apply the drag method, first the initial and final configuration are relaxed without any constraint for the hopping atom.
The relaxation yields, besides the binding energy and the relaxed hopping atom position ($E_\mathrm b^{\mathrm {i}}$ or $E_\mathrm b^{\mathrm {f}}$ and $\xvector{i}'$ or $\xvector{f}'$, respectively), three different outcomes:  a \textit{stable}, an \textit{unstable}, or a \textit{forbidden} configuration.
If an atom different from the hopping one moved too far  ($>0.4~r_0$, $r_0$ being the lattice constant), the configuration is regarded as forbidden.
If only the hopping atom moved too far, the configuration is regarded as unstable.
Otherwise, if no atom moved too far, it is regarded as stable.
We immediately rule out transitions between unstable configurations and transitions to forbidden ones. 
For the remaining transitions, the end points of the drag interval are determined as follows: For an unstable configuration, either $\xvector{i}$ or $\xvector{f}$ is an unstable site for which we assume the ideal fcc position; for the stable case, $\xvector{i}$ and/or $\xvector{f}$ are the projections of the relaxed positions $\xvector{i}'$ and/or $\xvector{f}'$ onto the hop direction.
Then, the diffusion path is first scanned with a relative coarse step width of $0.05~r_0$.
Afterwards a finer step width of $0.01~r_0$ is used to resolve the vicinity of the saddle point (see Fig.~\ref{fig:Examples}).
In the case of an unstable configuration, the saddle point may not be found within the drag interval, in which case the latter is extended beyond $\xvector{i}$ or $\xvector{f}$.
If the saddle point is still not found or if its distance from the ideal fcc site is $>0.4~r_0$, the transition and its reverse are ruled out as well.
Otherwise, the activation energy for the transition to the unstable configuration is found, while minus infinity is assigned to its reverse, i.e.\ the latter takes place immediately.

The presented calculation scheme is able to calculate such a pair of activation energies within 1--10 seconds on a 2.26 GHz Intel Core 2 Duo.
In the next section a method is presented to reduce the number of activation energies calculated on the fly.

\section{Activation energy pre-calculation\label{sec:precalculation}}
The vast number of possible configurations ($3\times 2^{72}+ 2^{79}\approx 6\times 10^{23}$) make a calculation of every activation energy in advance impossible.
But starting with an empty database, most of the early computing time would be spent on the calculation of new activation energies.
As shown below, the average number (averaged over $10^6$ KMC steps) of new processes per KMC step can become of the order of $10^{-1}$--$10^{-2}$ and thus the total number of processes can rapidly reach the order of $10^5$.
Fortunately, the rate of encountering new processes decreases fast enough, that an efficient simulation becomes possible at later simulation stages.
For comparison, the number of KMC steps per second would be of the order $10^4$ if no activation energy is calculated on the fly.
Because the KMC method is inherently serial, days would pass before the simulated system would make significant progress.
To circumvent this, transition configurations are gathered, using our recently presented three dimensional atomistic KMC simulation model\cite{Latz2012, Latz2012MRS}, which utilizes a simpler activation energy model.
In the simpler model the saddle point energies are not calculated, but constructed:
For hops to sites with $Z>2$, a constant model parameter $\Delta E$ is used instead and sites with $Z=1$ and 2 are regarded as unstable
\begin{equation}
E_{\mathrm{a}}=\left\{\begin{array}{l@{\ : \ }l}
					\max(E_{\mathrm{b}}^{\mathrm {f}} - E_{\mathrm{b}}^{\mathrm {i}},0)+\Delta E & Z>2\\
					E_{\mathrm{b}}^{\mathrm {f}} - E_{\mathrm{b}}^{\mathrm {i}} & Z\leq 2
				\end{array}\right.
\label{eq:Activationenergy}
\end{equation}

The activation energies of the gathered configurations can afterwards be calculated, as described in the previous section, using the drag method.
Since the activation energy calculations are independent of each other, they can be perfectly distributed on multiple processors (\textit{task farming}).
If the simplified model is a sufficient approximation, most of the transition configurations appearing with the new model will already be calculated.
In addition, with every simulation the number of already calculated activation energies increases.

In the next section, two examples are given which illustrate the applicability and performance of our model.

\section{Simulation\label{sec:Simulation}}
Two examples are presented to show the capabilities of our model.
First the diffusion of monolayer islands on Ag(111) is investigated.
Even in this example, mass transport between surface layers occurs occasionally.
As a real three dimensional example, the homoepitaxial growth of Ag on a Ag(111) surface is investigated at low temperatures.

\subsection{Diffusion of monolayer islands on Ag(111)}
In this section, the diffusion of monolayer islands on a Ag(111) surface is investigated.
The diffusion of single atoms also leads to the diffusion of the whole island.
The diffusion constant $D$ for large islands depends on the island radius $R$ as 
\begin{equation}
D \propto R^{-\alpha}\exp\left( - \frac{E_{\mathrm{eff}}}{k_\mathrm{B}T} \right)\, ,
\label{eq:D_size_dependence}
\end{equation}
where $E_{\mathrm{eff}}$ denotes an effective energy barrier and $\alpha$ is a scaling exponent\cite{Khare1995, Khare1996, Voter1986}.
$\alpha$ depends on the basic diffusion mechanism of the island.
Khare \textit{et al.}\cite{Khare1995, Khare1996} have shown that the diffusion of large clusters is a direct consequence of the fluctuations of the cluster boundary.
For three limiting cases, namely exclusive periphery diffusion, terrace diffusion, or evaporation-condensation, they found different scaling exponents of 3, 2, or 1, respectively.

Recently, the SLKMC model by Trushin et al.\cite{Trushin2005} was used to identify the influence of collective motion and multiple atom processes on the island diffusion\cite{Karim2011}.
They found, for increasing island size, a crossover from collective atom motion to periphery atom diffusion for Cu islands on Cu(111).
For large islands ($>19$ atoms), periphery diffusion by single-atom processes turned out to be the dominating mechanism.

The diffusion of large islands provides a perfect first example to test our simulation model.
To investigate the size dependence of the diffusion constant, the diffusion coefficients for 6 different island radii, 3.5, 5, 7, 10, 15 and $20~r_0$ (45--1452 atoms), at three different temperatures, 300, 500 and 700~K, were calculated.
The initial simulation setup consists of a $100\times100~r_0^2$ large, immobile Ag(111) substrate and an initially round island on top of it.
Periodic boundary conditions are used.

To reduce the number of activation energies calculated on the fly, an initial database is build up, using our recently presented three dimensional atomistic KMC simulation model\cite{Latz2012, Latz2012MRS}.
One preparatory simulation with the largest island radius ($R=20~r_0$) at an even higher temperature (800~K) was performed to gather occurring configurations within $10^9$ KMC steps.
In total, 406\,445 configurations were found, for which activation energies were calculated with the drag method and included in the initial database.
Since the calculation of the activation energy involves the corresponding back-process (cf.\ section \ref{sec:AEC}), 316\,754 additional configurations were obtained as a ``by-product'' and stored as well. 

The calculation took 280 CPU hours on Intel Xeon E5645 CPUs with 2.40~GHz, and were carried out perfectly in parallel.

\begin{figure}[tb]
\centering
\includegraphics[width=0.5 \linewidth]{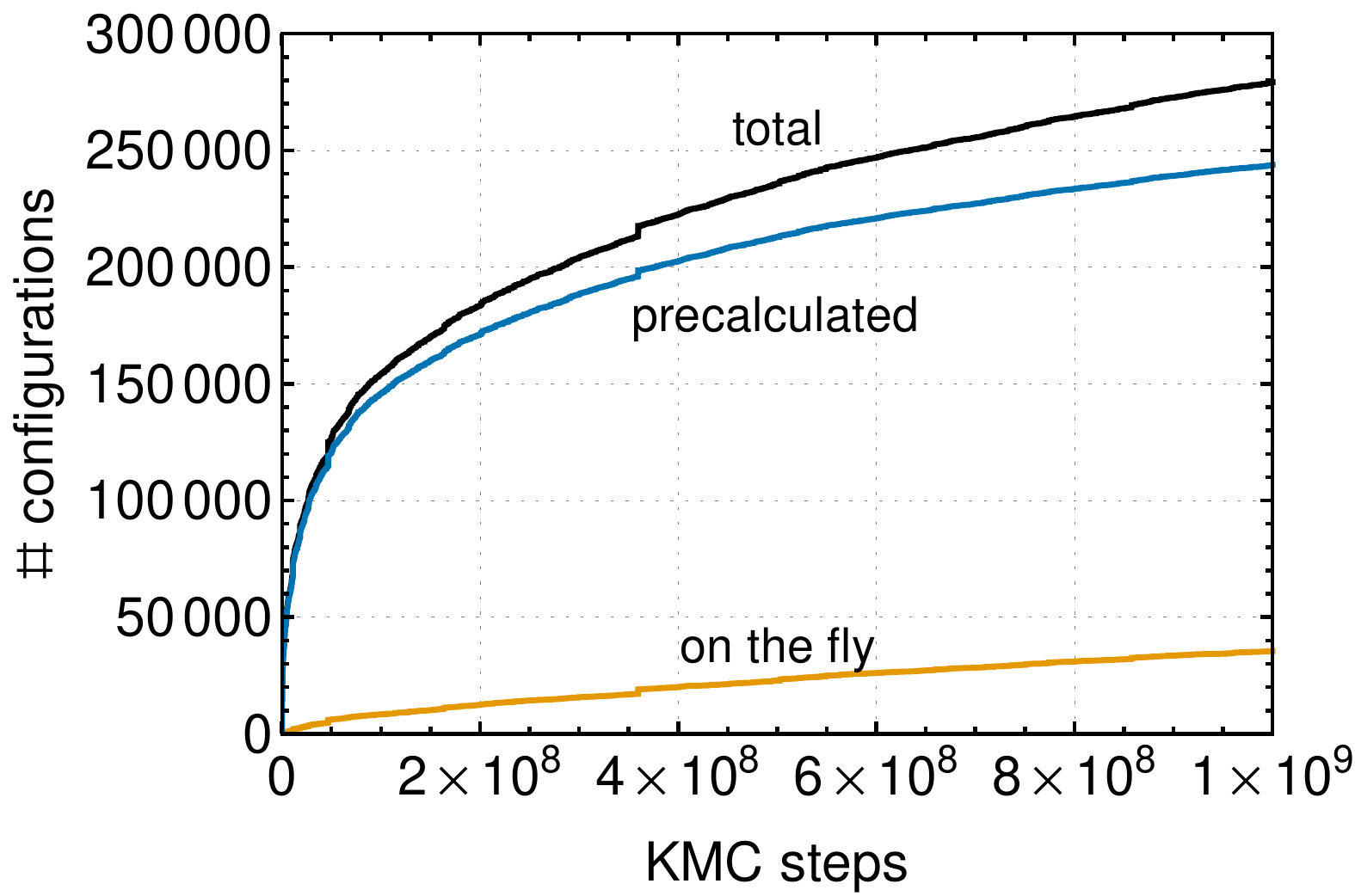}
\caption{
Accumulated number of configurations: calculated on the fly, precalculated, and their total for $R=20~r_0$, $T=700~\mathrm{K}$.
\label{fig:N_t}}
\end{figure}

To illustrate the impact of the initial database, the number of encountered configurations in a simulation of an island with $R=20~r_0$ and $T=700~\mathrm{K}$ is separated into configurations having been precalculated and the ones needed to be added on the fly (see Fig.~\ref{fig:N_t}).
It can be seen that especially at the beginning of the simulation, where the number of new configurations per KMC step is high, only few of them need to be calculated on the fly.

\begin{figure}[tb]
\centering
\includegraphics[width=0.5 \linewidth]{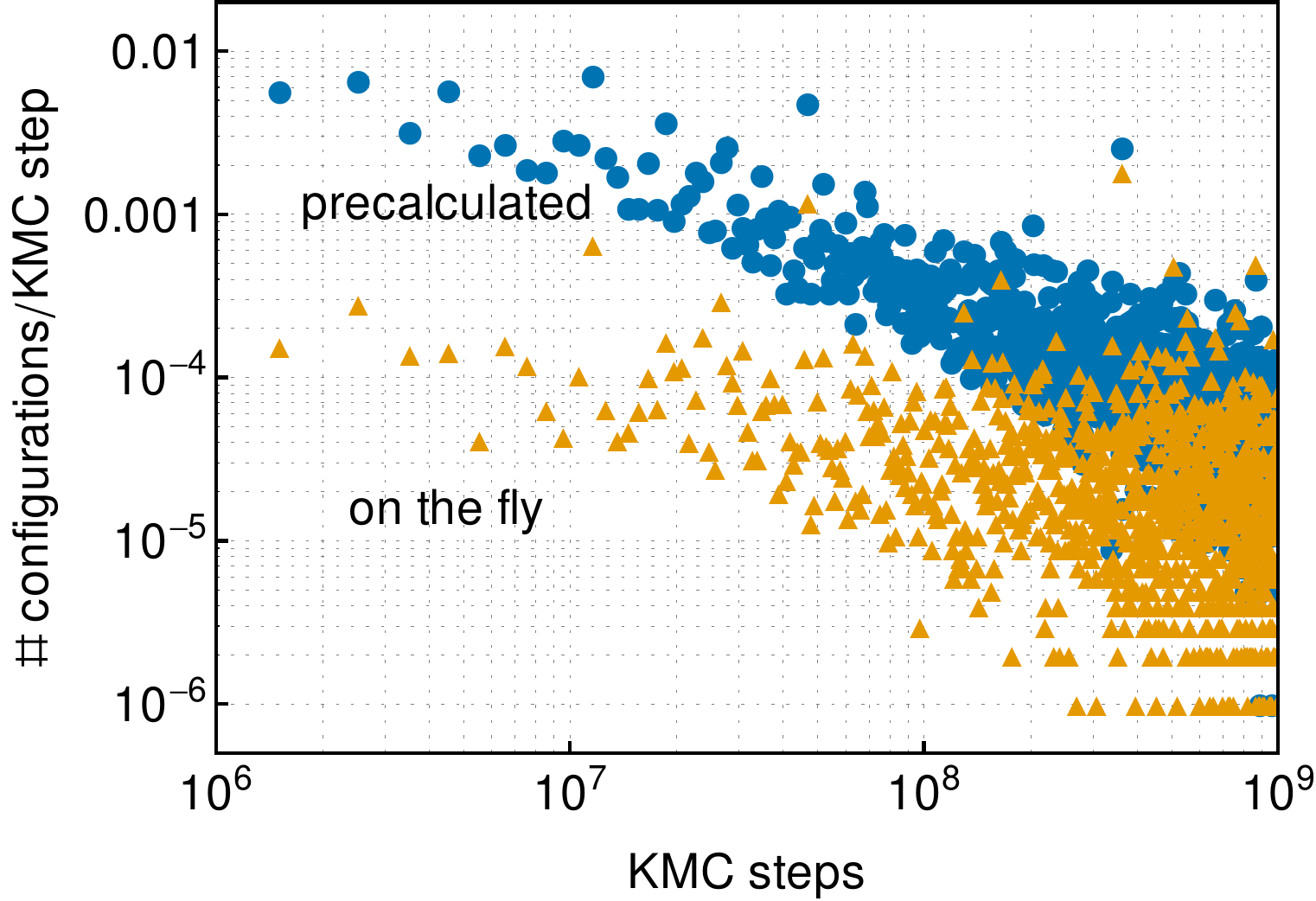}
\caption{Number of newly encountered configurations per KMC step (averaged over $10^6$ KMC steps), taken from the database and calculated on the fly for $R=20~r_0$, $T=700~\mathrm{K}$.
\label{fig:dN_t}}
\end{figure}

Fig.~\ref{fig:dN_t} shows the number of configurations per KMC step, which are encountered for the first time in a simulation run.
At the beginning, it goes up to approximately 0.01, but decreases roughly as a power law.

\begin{figure}[tb]
\centering
\includegraphics[width=0.5 \linewidth]{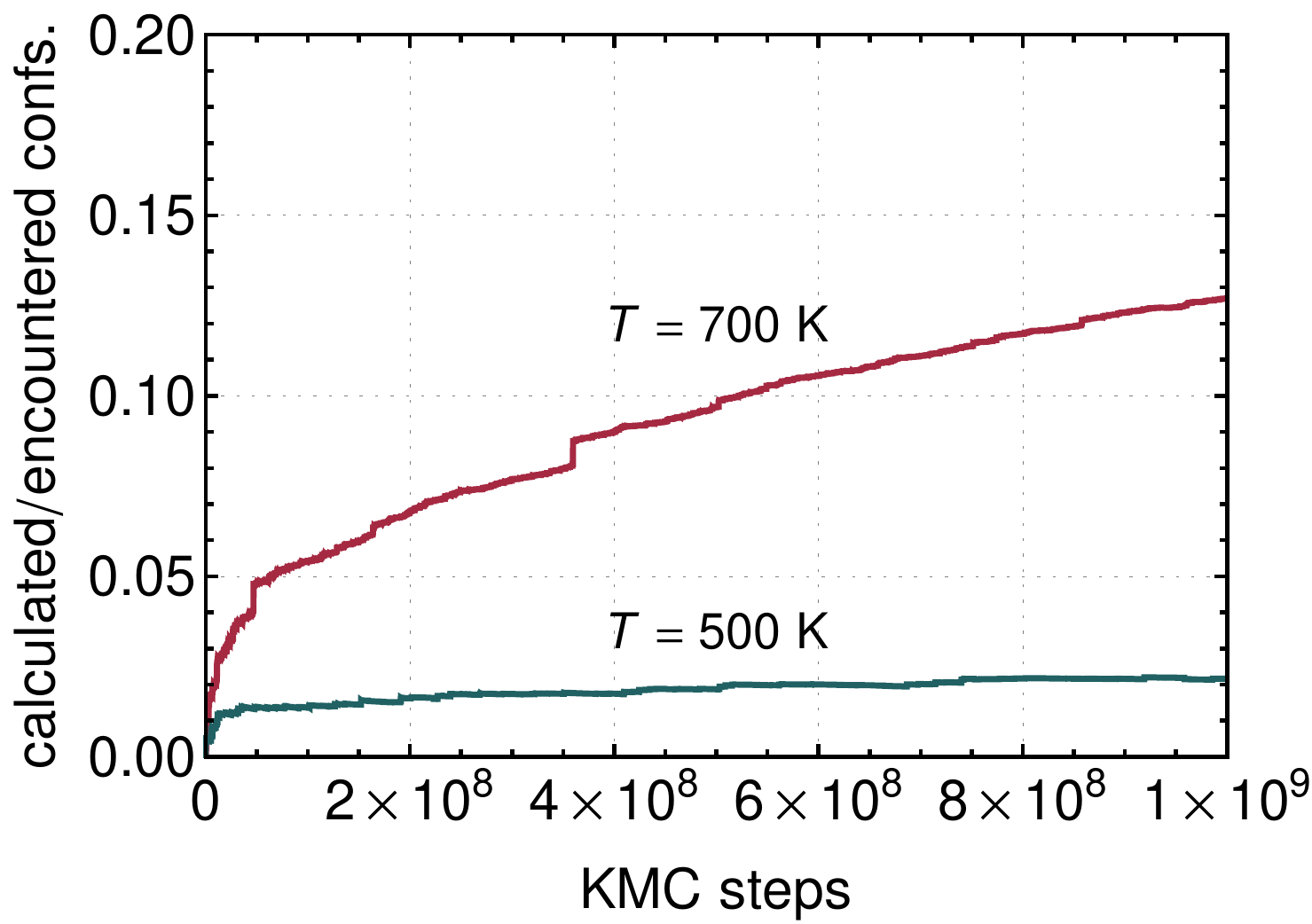}
\caption{
Ratio between occurring and on the fly calculated configurations ($R=20~r_0$).
\label{fig:ratio_N_t}}
\end{figure}

For better comparison, the ratio of configurations calculated on the fly to the the total number of newly encountered ones is shown in figure~\ref{fig:ratio_N_t}.
The ratio is slowly increasing up to approximately 13 \% for 700~K.
This increase is inevitable, but considering the fact that after $10^9$ KMC steps the number of encountered configurations amounts to 2/3 of the encountered ones in the preparatory simulation used for the initial database, 13~\% is relatively small.
If a smaller temperature is used, the ratio becomes even smaller, as can be seen in figure~\ref{fig:ratio_N_t} for $T=500$~K.
For $T=300~\mathrm{K}$ no activation energies needed to be calculated on the fly, because the dynamics was completely dominated by edge diffusion and the islands were strongly facetted.

In summary, the parallel pre-calculation of activation energies is a powerful way to significantly decrease the number of on the fly calculations and thus increase the simulation performance.
Especially at the beginning of a simulation, where the system would otherwise make only slow progress, a huge speed up is possible.
At later times, the rate of new configurations becomes small enough that a negative influence on the system performance is negligible.
During the simulations presented in this section, 753\,376 additional activation energies were calculated in total.
Thus, almost 50~\% of our database could be calculated in advance.
The achieved average program speed ranged, depending on the system size and temperature, between approximately 7\,000 and 30\,000 KMC steps per second on an Intel Xeon E5645 CPU with 2.40~GHz.
This is of the same order, which is achievable with simpler models\cite{Latz2012, Latz2012MRS, Zinetullin2010}.

The diffusion coefficient for two dimensional islands can be calculated as
\begin{equation}
D = {\frac{1}{4\Delta t}\langle\left| \vec{r}_{\mathrm{CM}}(t)-\vec{r}_{\mathrm{CM}}(t+\Delta t) \right|^2 \rangle}\, .
\label{eq:D_coefficient}
\end{equation}
$\vec{r}_{\mathrm{CM}}$ denotes the island‘s center of mass.
We calculated $D$ as a temporal average and chose $\Delta t = n\tau$, where $n$ is an integer and the time interval $\tau$ is chosen long enough for each temperature and island radius that it contains approximately $10^4$--$10^5$ steps.
Then, the diffusion coefficient can be extracted from a linear fit of $\left| \vec{r}_{\mathrm{CM}}(0)-\vec{r}_{\mathrm{CM}}(\Delta t) \right|^2$ as a function of the different $\Delta t$.

\begin{figure}[tb]
\centering
\includegraphics[width=0.5 \linewidth]{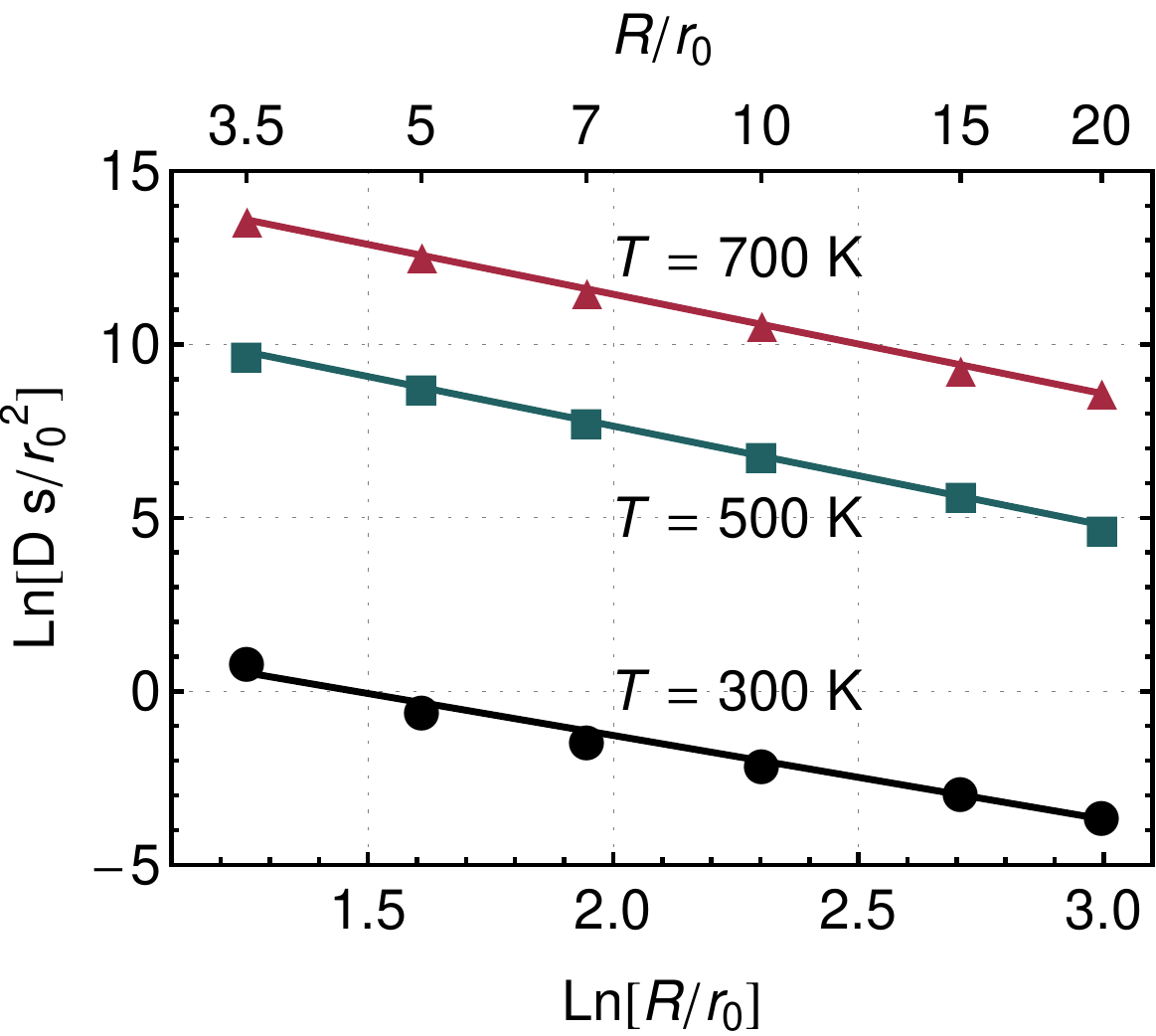}
\caption{Double logarithmic plot of the monolayer island diffusion coefficient for different island radii.
\label{fig:D}}
\end{figure}

The measured diffusion constants are shown in figure~\ref{fig:D}.
As predicted, the diffusion coefficient decreases with increasing island radius, according to a power law.

The scaling exponents are:
\begin{eqnarray*}
\alpha(T=300~\mathrm{K})&\approx 2.42 \pm 0.57,\\
\alpha(T=500~\mathrm{K})&\approx 2.86\pm 0.05,\\
\alpha(T=700~\mathrm{K})&\approx 2.87\pm 0.05.
\end{eqnarray*}
This corresponds to a dominating periphery diffusion ($\alpha=3$) for Ag(111) islands with $R\in [3.5~r_0,20~r_0]$.
The deviations to smaller $\alpha$ can be understood due to an increasing contribution of terrace diffusion with increasing island size.
For large islands a crossover from periphery to terrace diffusion is expected\cite{Ala-Nissila2002}.
In our simulations, terrace diffusion via adatoms on the substrate and vacancies within the island could be observed with increasing temperature and island size.

A mass transport by adatoms hopping on top of the island and reattaching at a different island edge position appears only very rarely.

\subsection{Homoepitaxial film growth of Ag on Ag(111)}
Homoexpitaxial growth provides a well suited example to demonstrate the capabilities of our model by a real three dimensional problem.
In this section, the homoepitaxial growth of Ag on Ag(111) is investigated at low temperatures (150--300~K).

At the submonolayer stage, a distribution of two dimensional islands forms.
At later stages, multilayer stacks of 2D islands form due to the presence of a large Ehrlich-Schwoebel barrier\cite{Li2009}.

\begin{figure}[tb]
\centering
\includegraphics[width=0.4 \linewidth]{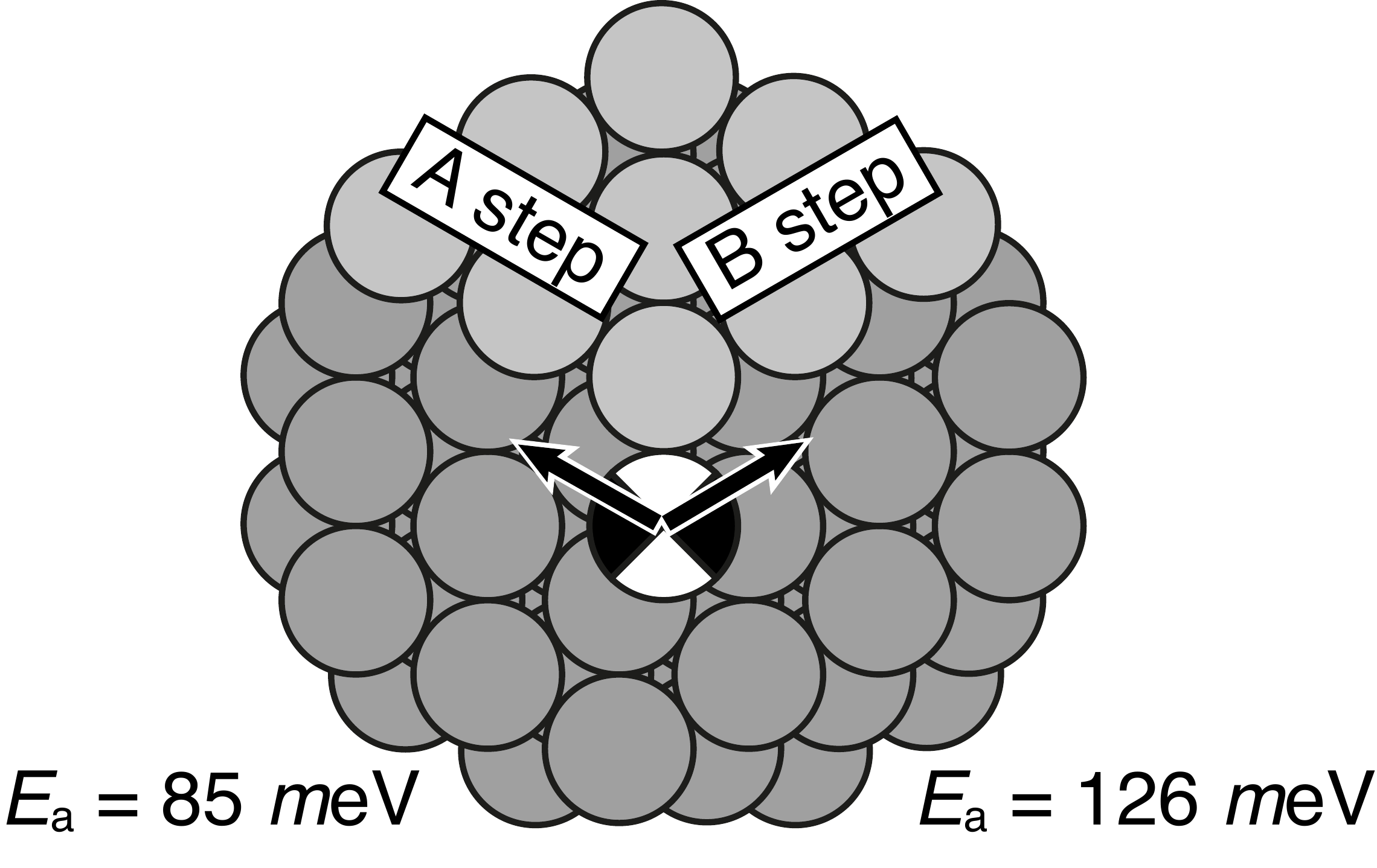}
\caption{Corner diffusion anisotropy.
\label{fig:CDA}}
\end{figure}

Islands grow due to the accumulation of deposited atoms.
Atoms, which aggregate at the corner of an island (Fig.~\ref{fig:CDA}) have a notably smaller activation energy to reach A-steps than B-steps:
The activation energies in our model are 85~meV and 126~meV, respectively.
The values are close to those obtained previously using an embedded atom potential\cite{Cox2005}.
This corner diffusion anisotropy (CDA) leads at low temperatures to dendritic islands, which follow the threefold symmetry of the underlying substrate\cite{Cox2005}.
Atoms aggregating at the corner of an island attach more frequently at A steps than B steps, leading to an anisotropic island growth in the direction of the A steps.
With increasing temperature, periphery diffusion becomes fast enough that the shape of the islands gets more compact.

Studies have shown that cluster diffusion cannot significantly affect the submonolayer and post deposition coarsening of submonolayer island distributions, as well as the multilayer growth, see \cite{Li2009} for a detailed discussion.
Hence, we can neglect collective processes also in this scenario.

To investigate the different growth regimes, our initial simulation setup consists of three $50\times50~\mathrm{nm}^2$ ($202\times 176$ atoms) large immobile atom layers.
New atoms are randomly deposited on the surface with the rate $F$ (in monolayers (ML) per second).
If the deposition of an atom leads to an configuration that could not be generated by diffusion processes, the atom is placed at an allowed nearest neighbor site instead.

Besides the number of activation energies calculated on the fly, the simulation performance depends sensitively on the ratio between the adatom hopping rate and the deposition rate of new atoms.
The hopping rate for adatoms on an fcc (111) surface is very high due to the small activation energy ($67~\mathrm{meV}$).
Since at later simulation stages freely diffusing adatoms on islands are almost always present, the average KMC time step becomes of the order of the inverse adatom hopping rate ($10^{-10}~\mathrm{s}$ for $T=180~\mathrm{K}$).
Typical experimental values for $F$ are of the order $10^{-3}~\mathrm{ML/s}$, which corresponds to approximately 0.03~s between atom depositions for our  system size.
Thus on average $10^8$--$10^9$ KMC steps are performed before the next atom is deposited.
This is beyond the capabilities of KMC simulations.
Therefore, a vastly increased deposition rate ($F=10~\mathrm{ML/s}$) was used.
The increased deposition rate results in an increased island density\cite{Venables1984, Schroeder1995}.

For the homoepitaxy simulations, the database from the diffusion simulations of the previous section was used.
No additional activation energies were precalculated, since the simplified activation energy model is a rather poor approximation at low temperatures.
But simulations running in parallel used the same database and were thus able to benefit from each others activation energy calculations.

\begin{figure}[tb]
\centering
\includegraphics[width=0.5 \linewidth]{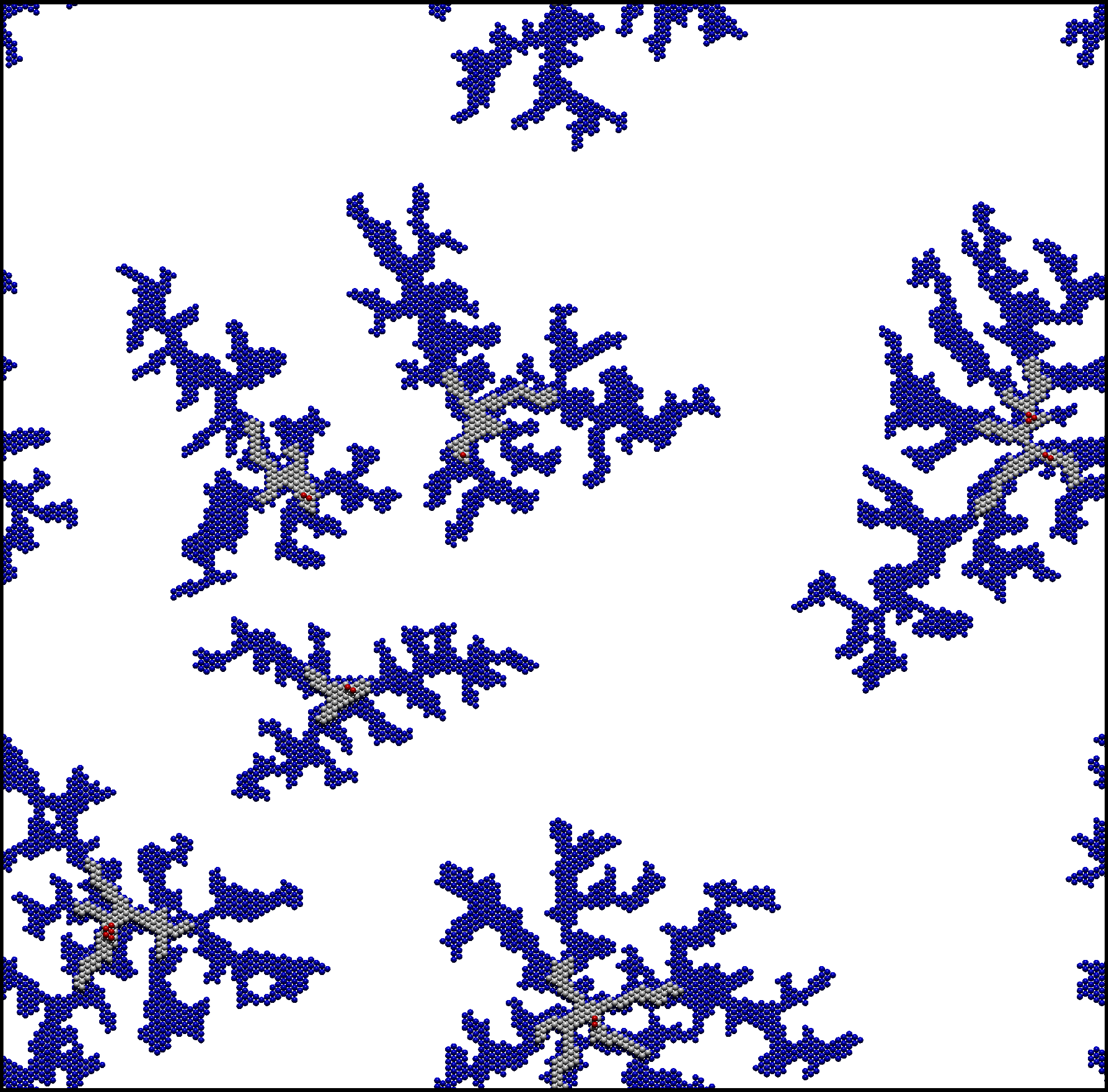}
\caption{Dendritic islands at $T=150~\mathrm{K}$ after deposition of 0.2~ML onto a $50\times50~\mathrm{nm}^2$ Ag(111) surface.
\label{fig:homoepitaxy150}}
\end{figure}

To investigate the different growth regimes, simulations were performed for $T=150$~K, 230~K and 300~K.
In figure~\ref{fig:homoepitaxy150}, the system is shown for $T=150~\mathrm{K}$ after deposition of 0.2~ML (approximately $1.8\times10^9$ KMC steps).
Since the structure is rather fractal and, in contrast to the diffusion simulation, three dimensional, the number of activation energies calculated on the fly, is comparatively big (approximately 135\,000).
The achieved KMC steps per second still range between $10^3$ and $10^4$.
The growth proceeds predominantly in the direction of the A steps, leading to dendritic islands.
The second layer islands are located at the center of the islands in the first layer, where a stable nucleus formed.
In contrast to experiments\cite{Evans2006,Li2009} the regime, where dendritic islands are present, extends from 120--135~K to slightly higher temperatures, because the periphery diffusion is not fast enough to compensate the increased growth rate.

\begin{figure}[tb]
\centering
\includegraphics[width=0.5 \linewidth]{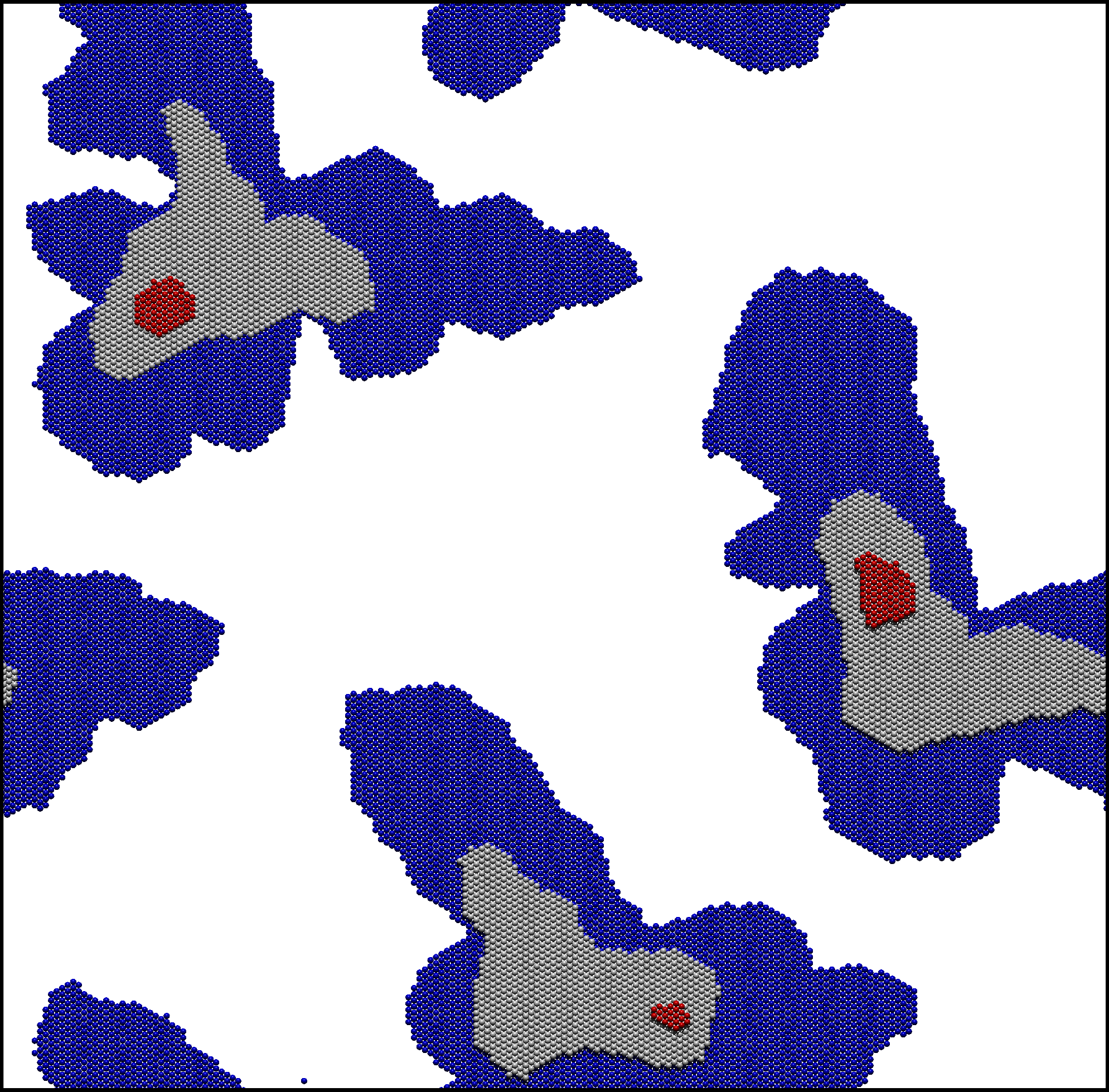}
\caption{Compact islands at $T=230~\mathrm{K}$ after deposition of 0.4~ML onto a $50\times50~\mathrm{nm}^2$ Ag(111) surface.
\label{fig:homoepitaxy230}}
\end{figure}

\begin{figure}[tb]
\centering
\includegraphics[width=0.5 \linewidth]{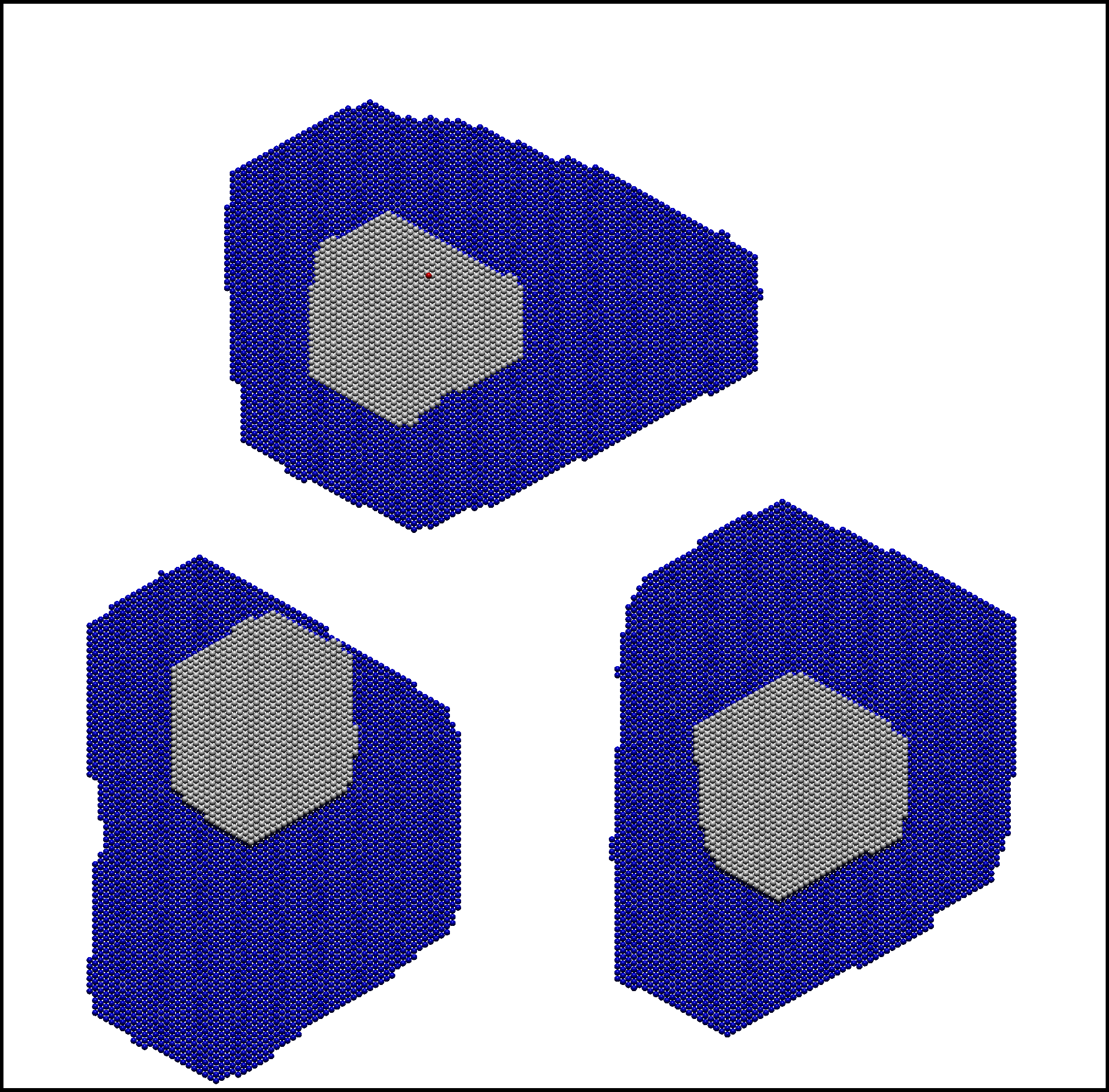}
\caption{Hexagonal islands at $T=300~\mathrm{K}$ after deposition of 0.4~ML  onto a $50\times50~\mathrm{nm}^2$ Ag(111) surface.
\label{fig:homoepitaxy300}}
\end{figure}

With increasing temperature, the periphery diffusion becomes fast enough that the shape of the islands gets more compact.
Figure~\ref{fig:homoepitaxy230} and \ref{fig:homoepitaxy300} show the system after deposition of 0.4~ML at $T=230~\mathrm{K}$ (approximately $2.3\times10^9$ KMC steps) and $T=300~\mathrm{K}$ (approximately $1.4\times10^9$ KMC steps), respectively.
Due to the less fractal shape the number of activation energies calculated on the fly reduced to approximately 41\,000 and 7\,000, respectively.
At $T=300~\mathrm{K}$, the temperature is already high enough for hexagonal islands to form, a shape closer to the equilibrium one.

\section{Conclusions\label{sec:Conclusion}}
The accuracy of KMC simulations depends crucially on a correct rate catalog.
We presented a model that is able to calculate on the fly configuration dependent atom diffusion rates for three dimensional systems and reuse them via a pattern recognition scheme.
The number of local configurations occurring during a simulation can be tremendous for three dimensional systems, which makes an efficient simulation difficult.
We showed that by setting up an initial database, the number of rates calculated on the fly can be decreased significantly.
Moreover, while parallel KMC simulations in general are highly problematic\cite{Shim2005, Shi2007}, the pre-calculation can be perfectly parallelized, which pays off for SLKMC simulations, where the calculation of the energy barriers dominates the CPU usage.
In a similar manner, SLKMC simulations running in parallel can communicate their calculated activation energies with each other.
Furthermore, since the database is saved it gets more and more complete with each simulation having been run, increasing the performance of subsequent simulations.
In the case of changing the interaction potential (e.g.\ for a different material), the whole database can be recalculated in parallel.
Two simulation examples are presented that show the versatility and performance of our model.
Without any adjustment of model parameters, the different island growth shapes observed for homoepitaxial film growth of Ag on Ag(111) at different temperatures agree with experiments.
This is remarkable, as for example dimer diffusion was not taken into account, although its rate is comparable to monomer diffusion on Ag(111) \cite{Nandipati2012}.
Such additional processes could improve the results quantitatively at the expense of a more time consuming simulation.

Despite the on the fly calculation of new rates, simulations for systems consisting of up to several thousand atoms were possible for up to $10^{10}$ KMC steps.
The number of achievable KMC steps per second is thus not much lower than for simpler approaches\cite{Latz2012, Latz2012MRS}.

The diffusion of monolayer islands on Ag(111) revealed a dominating periphery diffusion for islands between 3.5 and 20~$r_0$, and temperatures between 300 and 700~K.
For the homoepitaxial film growth of Ag on Ag(111) at low temperatures two different growth regimes could be reproduced.

In the future, our KMC model could be expanded by including multi-atom and concerted moves, which are needed to correctly implement the diffusion processes for small clusters at moderate temperatures. 
For example, the test system considered in this work would benefit from the inclusion of the dimer diffusion on the (111) surface. Also the direct consideration of hcp sites is possible in principle.
But it has to be kept in mind that such extensions, while being rather straight forward in principle, increase the amount of possible configurations and transitions tremendously, which can have a devastating impact on the performance.

\ack
Financial support from the Deutsche Forschungsgemeinschaft through SFB616 ``Energy Dissipation at Surfaces'' is gratefully acknowledged.

\bibliographystyle{unsrt} 
\section*{References}
\bibliography{Paper}

\end{document}